\documentclass[lettersize,journal]{IEEEtran}
\usepackage{cite}
\usepackage{amsmath,amssymb,amsfonts}
\usepackage{algorithmic}
\usepackage{graphicx}
\usepackage{subcaption}
\usepackage{textcomp}
\usepackage{multirow}
\usepackage{balance}
\usepackage{color,soul}
\usepackage{titlesec}
\usepackage[left=1.41cm,right=1.41cm,top=1.8cm,bottom=4.2cm]{geometry}

\def\BibTeX{{\rm B\kern-.05em{\sc i\kern-.025em b}\kern-.08em
    T\kern-.1667em\lower.7ex\hbox{E}\kern-.125emX}}
\begin{document}
\pagenumbering{gobble} 
\title{Super-Directive Antenna Arrays: How Many Elements Do We Need?} 
\author{\IEEEauthorblockN{
Ihsan Kanbaz\IEEEauthorrefmark{1}\IEEEauthorrefmark{2},   
Okan Yurduseven\IEEEauthorrefmark{1},   
and Michail Matthaiou\IEEEauthorrefmark{1}   
}                                     
\\
\IEEEauthorblockA{\IEEEauthorrefmark{1}
Centre for Wireless Innovation (CWI), Queen’s University Belfast, Belfast BT3 9DT, U.K. }
\\\IEEEauthorblockA{\IEEEauthorrefmark{2}
Department of Electrical and Electronics Engineering, Gazi University, Ankara, Turkey 
\\e-mail: \{i.kanbaz, okan.yurduseven, m.matthaiou\}@qub.ac.uk}
}
\maketitle

\begin{abstract} 
Super-directive antenna arrays have faced challenges in achieving high realized gains ever since their introduction in the academic literature. The primary challenges are high impedance mismatches and resistive losses, which become increasingly more dominant as the number of elements increases. Consequently, a critical limitation arises in determining the maximum number of elements that should be utilized to achieve super-directivity, particularly within dense array configurations. This paper addresses precisely this issue through an optimization study to design a super-directive antenna array with a maximum number of elements. An iterative approach is employed to increase the array of elements while sustaining a satisfactory realized gain using the differential evolution (DE) algorithm. Thus, it is observed that super-directivity can be obtained in an array with a maximum of five elements. Our results indicate that the obtained unit array has a $67.20\%$ higher realized gain than a uniform linear array with conventional excitation. For these reasons, these results make the proposed architecture a strong candidate for applications that require densely packed arrays, particularly in the context of massive multiple-input multiple-output (MIMO).
\let\thefootnote\relax\footnotetext{This work was supported by a research grant from the European Research Council (ERC) under the European Union’s Horizon 2020 research and
innovation programme (grant No. 101001331).}
\end{abstract}


\maketitle
\section{Introduction}
The increasing interference in the frequency spectrum and the growing demand for faster data transmission have prompted researchers to focus on advanced communication systems at higher frequencies, such as 5G and beyond \cite{6G,6G_2,Prospective}. Due to the shorter wavelengths (higher frequency), it becomes possible to pack an unconventionally high number of antennas within finite volumes in order to compensate the exacerbated path loss and penetration losses. Not surprisingly, there has been a substantial surge in research efforts to develop cost-effective yet efficient hardware units, primarily focusing on antenna arrays \cite{6G_3}.

Increasing the number of antennas within a confined space magnifies the mutual coupling effect in antenna arrays, degrading the array's performance and introducing complex mathematical challenges for accurately calculating the current distributions of the array elements. Consequently, antenna engineers have focused on mitigating this mutual coupling effect and developing various array synthesis techniques. However, adapting these techniques to antenna array configurations with distinct geometric shapes proves to be a complex task. Despite this, a widely accepted approach for reducing the mutual coupling effect across all geometric shapes and array types involves maintaining an inter-element distance greater than half a wavelength \cite{CouplingMatrix}. Nevertheless, this method leads to an increase in the antenna size. This increase in the array form factor can be a significant limitation for massive MIMO applications, where arrays with numerous antenna terminals are deployed \cite{MM-wave}. Therefore, recent attention has been directed towards the array configurations with inter-element-antenna distances below half a wavelength, seeking for a more suitable solution \cite{HASKOU,Altshuler,Marzetta2}.

Contrary to the common belief, a historically seminal study conducted by Uzkov uncovered an unexpected insight: reducing the distance between the isotropic antennas can improve the performance of an antenna array \cite{uzkov}. The traditional array systems typically assume that the directivity of an array increases linearly with the number of elements, but Uzkov demonstrated an intriguing alternative. He articulated that as the antenna spacing approaches zero in an ideal scenario, the resulting array directivity becomes proportional to the square of the number of elements. This shift in thinking has given rise to the concept of 'super-directive arrays' where the focus is on creating more compact antenna arrays to achieve higher directivity \cite{Altshuler,SuperDirectiveAntennas}.

It is commonly acknowledged that there is a strong relationship between the directivity and the gain in antenna systems. The calculation of the gain includes both the directivity and radiation efficiency, as explained by Balanis \cite{balanis}. In the context of super-directive arrays, where the challenges pertaining to the radiation efficiency are fundamental, super-conducting materials are suggested to address this issue. Another facet that merits consideration is related with the expression for the realized gain, which encompasses not only the radiation efficiency but also considers the line impedance mismatch \cite{Limit2}. 

The overall efficiency, combining the feedline and radiation efficiency, can be notably high in an ideally matched array. However, the scenario shifts when dealing with super-directive arrays, where the realized gain experiences a notable reduction due to the line mismatch occurring among certain array elements. For this reason, there has been some intensive work on increasing the realized gain in super-directive arrays \cite{MultiOpt,OntheMax,Tornesse, Eucap}. Although these studies, involving complex optimization and geometric alignment techniques, provide valuable results for arrays with a few elements, achieving high realized gains becomes increasingly difficult as the number of elements increases. Most importantly, the maximum number of elements at which the maximum high realized gains can be achieved remains an open question.

Therefore, in this paper, we aim to determine the parameters such as element spacing, antenna feed currents, and antenna dimensions for arrays with the highest number of antennas obtained through multi-parameter optimization to achieve high realized gain. We then analyze this array using full-wave electromagnetic simulation software. Finally, we conclude that the proposed array, with its high realized gain, is suitable for dense array architectures.

The paper is organized as follows: Section \ref{sec:SuperDirectivity} introduces the theoretical framework of super-directive arrays. Section \ref{sec:Opt} delves into the multi-parameter optimization for super directivity, employing the DE algorithm to optimize a dipole array with a maximum number of elements. In Section \ref{sec:Num}, numerical results and comparisons for the maximum element dipole array are presented. Finally, Section \ref{sec:Conclusion} provides a conclusion, summarizing key findings and highlighting the potential of the proposed unit array in the dense array scenarios, particularly in massive MIMO applications.

\textit{Notations}: The bold letters stand for vectors and matrices; $\textbf{X}^T$ $\textbf{X}^*$ and $\textbf{X}^H$ represent the transpose, conjugate and conjugate transpose operations, respectively. Moreover, $\|{\bf{x}}\|^2 $ denotes the $l_2$ norm of  $\textbf{x}$.
\begin{figure}
    \centering
    \includegraphics[scale=0.75]{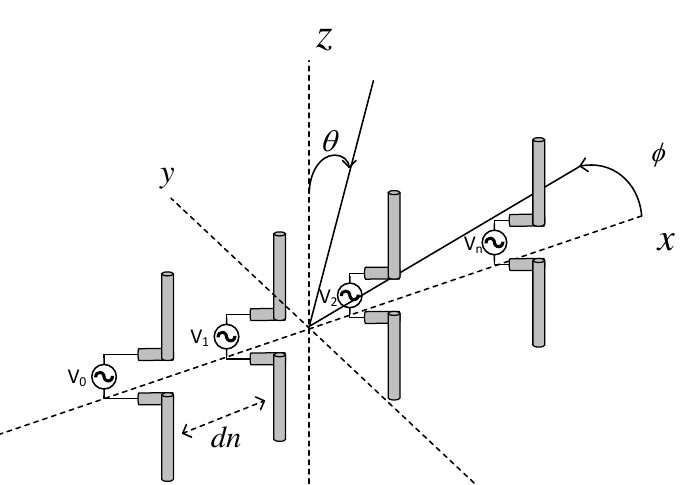}
    \caption{The geometry of dipole antenna array.}
    \label{Fig1}
\end{figure}
\section{Super Directivity on Antenna Arrays} 
\label{sec:SuperDirectivity}
Considering an array of $N$ dipole elements (see Fig. \ref{Fig1}), the current distribution of a single dipole with a length of $l$ located in the center of the array can be expressed as follows \cite[Eq.~(4-56)]{balanis}:
\begin{equation}
\mathbf{I}_{e}(z', x'=0, y'=0) = i_\mathrm{n}(0) \frac{\sin\left(k\left(\frac{l}{2}-|z'|\right)\right)}{\sin\left(\frac{kl}{2}\right)} \mathbf{a}_z, 
\label{eq1}
\end{equation}
for $ -l/2 \leqslant z' \leqslant l/2 $. Here, $k=2\pi/\lambda$ is the wave number, $\lambda=c/f$ is the wavelength, $i_n(0)$ is the complex excitation current, while $c$, $\mathbf{a}_z$ and $f$ represent the speed of light, the unit vector along the $z$ direction, and center frequency, respectively. For a far-field point, the electric field of each dipole is written as follows:
\begin{equation}
\mathbf{E}_\mathrm{n}(r,\theta) = -j\eta \mathbf{F}(\theta) \cdot \frac{e^{-jkr}}{2\pi r} \cdot i_\mathrm{n}(0). \label{eq2}
\end{equation}
Here, $\eta\approx120\pi$, $\theta$, and $r$ represent the impedance of free space, the observation angle, and the distance of the array element from the observation point, respectively. The term $\mathbf{F}(\theta)$ is the element factor and is theoretically expressed as follows \cite[Eq.~(4-62a)]{balanis}:
\begin{equation}
\mathbf{F}(\theta) = \frac{\cos\left(\frac{kl}{2}\cos\theta\right)-\cos\left(\frac{kl}{2}\right)}{\sin(\theta)}\textbf{a}_{{\theta}} \label{eq3}
\end{equation}
Here, $\textbf{a}_{{\theta}}$ represents the unit vector along the polar direction. The total electric field of the array can be expressed as follows:
\begin{equation}
\textbf{E}(r,\theta) = -j\eta \frac{e^{-jkr}}{2\pi r} \sum_{n=0}^{N-1}e^{-jk\mathbf{r}\cdot\mathbf{r}_\mathrm{n}}i_{n}(0)\mathbf{F}(\theta). \label{eq4}
\end{equation}
Here, $\textbf{r}\overset{\Delta}{=}(\sin\theta \cos\phi,\sin \theta \sin \phi, \cos \theta)^T$ is the position vector directed to the observation location, while $\textbf{r}_n\overset{\Delta}{=}(dn,0,0)$ represents the position vector of the $n$-th element. For an antenna whose electric field value is known, the radiation intensity can be written as follows \cite{dovelos}:
\begin{equation}
\begin{aligned}
U(\theta) = \frac{r^2}{2\eta}\|{\mathbf{E}}(r,\theta)\|^2 \
= \frac{\eta}{8\pi^2} \|{\mathbf{F}}(\theta)\|^2 \left|{\mathbf{a}^H(\theta) \cdot \mathbf{i}}\right|^2. \label{eq1}
\end{aligned}
\end{equation}
Here, ${\mathbf{a}}(\theta) = \begin{bmatrix} e^{-jk\mathbf{r}\cdot\mathbf{r}_0}, ... , e^{-jk\mathbf{r}\cdot\mathbf{r}_n} \end{bmatrix}^T \in \mathbb{C}^{N \times 1}$  and ${\mathbf{i}} = \begin{bmatrix} i_\mathrm{0}(0), ... , i_\mathrm{N-1}(0) \end{bmatrix}^T \in \mathbb{C}^{N \times 1}$ are the far-field array response vector and vector input currents, respectively. The total radiated power is obtained by integrating the radiation intensity as follows \cite{Orfanidis}:
\begin{equation}
\begin{split}
&P_\mathrm{rad} = \int_0^{2\pi} \int_0^{\pi} U(\theta) \sin\theta \mathrm{d}\theta \mathrm{d}\phi \\
&= \frac{\eta}{8\pi^2} {\bf{i}}^H \cdot \mathop{\underbrace{\int_0^{2\pi} \int_0^{\pi} {\bf a} (\theta) \cdot {\bf a}^H (\theta) \|{\bf{F}}(\theta)\|^2 \sin\theta \mathrm{d}\theta \mathrm{d}\phi}}_{\Re\{\bf{Z}\}'} \cdot {\bf{i}}\\
&=\frac{1}{2} {\bf{i}}^H \cdot \Re{\{\bf{Z}\}'} \cdot {\bf{i}},
\label{Prad}
\end{split}
\end{equation}
where, $\Re\{\cdot\}$ represents the real part operation, and $\mathbf{Z}' \in \mathbb{C}^{N \times N}$ represents the input impedance matrix of the lossless dipole array. 

The phenomenon of self-impedance is demonstrated when $m=n$ (expressed as $Z_{n,n}$), and it is referred to as mutual impedance (denoted as $Z_{m,n}$) when $m$ and $n$ are not equal. In a practical context, evaluating the conductor losses of the dipoles becomes imperative. The computation of loss resistance for a current-carrying conductor wire can be described as follows\cite{dovelos}:
\begin{equation}
R_\mathrm{loss}\overset{\Delta}=\bar{R}_\mathrm{loss} \int_{-l/2}^{l/2} {\left| \frac{I_\mathrm{e}}{i_n} \right|}^2 \mathrm{d}x' \\ = \frac{1}{4k\rho} \sqrt{\frac{f\mu}{\pi \sigma_\mathrm{c}}} \frac{kl-\sin(kl)}{\sin^2(kl/2)}, 
\label{eq:Rloss}
\end{equation}
where
\begin{equation}
		\bar{R}_\mathrm{loss}=\frac{1}{2\rho} \sqrt{\frac{f\mu}{\pi \sigma_c}}, \label{eq}
	\end{equation}
where $\rho=\lambda/200$ is the radius of the dipole, $\mu=4\pi\cdot10^{-7} \mathrm{H/m}$ is the permeability of free space, and $\sigma_c=5.8\cdot10^{7} \mathrm{S/m}$ is the conductivity of the copper. Therefore, the overall power loss is calculated as follows:
\begin{equation}
P_\mathrm{loss} = \frac{1}{2}R_\mathrm{loss}|{\mathbf{i}}|^2 = \frac{1}{2} {\mathbf{i}^H \cdot {\mathbf{R}_\mathrm{loss}} \cdot \mathbf{i}, \label{eq5}}
\end{equation}
where $\mathbf{R}_\mathrm{loss} = R_\mathrm{loss} \mathbf{I}_N \in \mathbb{R}^{N \times N}$, and $\mathbf{I}_N$ represents the $N \times N$ identity matrices.
    
    From this point of view, the total input power at the antenna feed point is calculated as follows:
    \begin{eqnarray}
    \begin{split}
&P_\mathrm{in} = P_\mathrm{loss} + P_\mathrm{rad}
\\ 
&= \frac{1}{2} {\bf{i}}^H \cdot {\bf R}_\mathrm{loss} \cdot {\bf i} + \frac{1}{2} {\bf{i}}^H \cdot \Re{\{\bf{Z}\}'} \cdot {\bf{i}} = \frac{1}{2} {\bf{i}}^H \cdot \Re{\{\bf{Z}\}} \cdot {\bf{i}} 
   \label{Pin}
   \end{split}
	\end{eqnarray}
	where
	\begin{equation}
		\Re{\{\bf{Z}\}} \overset{\Delta}= 
		\begin{bmatrix} 
			R_\mathrm{s} + R_\mathrm{loss} & R_\mathrm{m} \\
			R_\mathrm{m} & R_\mathrm{s} + R_\mathrm{loss} \\
		\end{bmatrix} \in \mathbb{R} ^{N \times N} .
  \label{ReZ}
	\end{equation}
In this context, the variables $R_\mathrm{s}$ and $R_\mathrm{m}$ represent the real parts of self and mutual impedances, respectively. Subsequently, the antenna array gain can be expressed in terms of the radiation intensity and input power as follows \cite{dovelos}:
 \begin{eqnarray}
\begin{split}
&G(\theta) = \frac{4\pi U(\theta)}{P_\mathrm{in}} = G_d(\theta)(R_\mathrm{loss}+R_\mathrm{s}) \frac{\left|{\bf a}^H (\theta) \cdot {\bf{i}}\right|^2}{{\bf{i}}^H \cdot \Re{\{\bf{Z}\}} \cdot {\bf{i}}} ,
\label{G1}
\end{split}		
\end{eqnarray}
 where 
 \begin{equation}
     G_\mathrm{d}(\theta)=\frac{\eta}{\pi (R_\mathrm{loss}+R_s) }\| {\bf{F}}(\theta)\|^2,
 \end{equation} 
 represents each gain of the isolated dipole. Here, it is assumed that the active element patterns of the dipoles do not change due to the arrangement of the antennas for ease of calculation.

 \section{Optimization Approach to Achieve the Maximum Number of Elements}
\label{sec:Opt}
In this section, an analysis of the maximum number of elements that can be optimized is conducted using a DE-based approach. The optimization process is initiated with two elements and arrays with up to ten elements are considered by evaluating the performance of arrays that can achieve the target realized gain. The optimization parameters, including the population size ($NP$) of $200$, a cross-over factor ($CR$) of $0.8$, a mutation factor ($F$) of $0.8$, and a maximum number of iterations of $250$, are carefully selected. The primary objective of this optimization is to minimize the mean squared error (MSE) between the desired realized gains (which are set at $40\%$ higher than the realized gain when mutual coupling is neglected) and the calculated realized gains. The following cost function quantifies this optimization objective:

\begin{equation}
F_\mathrm{cost} = H(\Delta|\mathrm{RG}) \Delta|\mathrm{RG}.
\label{eq:cost}
\end{equation}
Here, the Heaviside step function, denoted as $H(\cdot)$, is utilized, while $\Delta|\mathrm{RG}$ is defined as the MSE between the desired and calculated realized gains.
\begin{figure}
    \centering
    \includegraphics[scale=0.65]{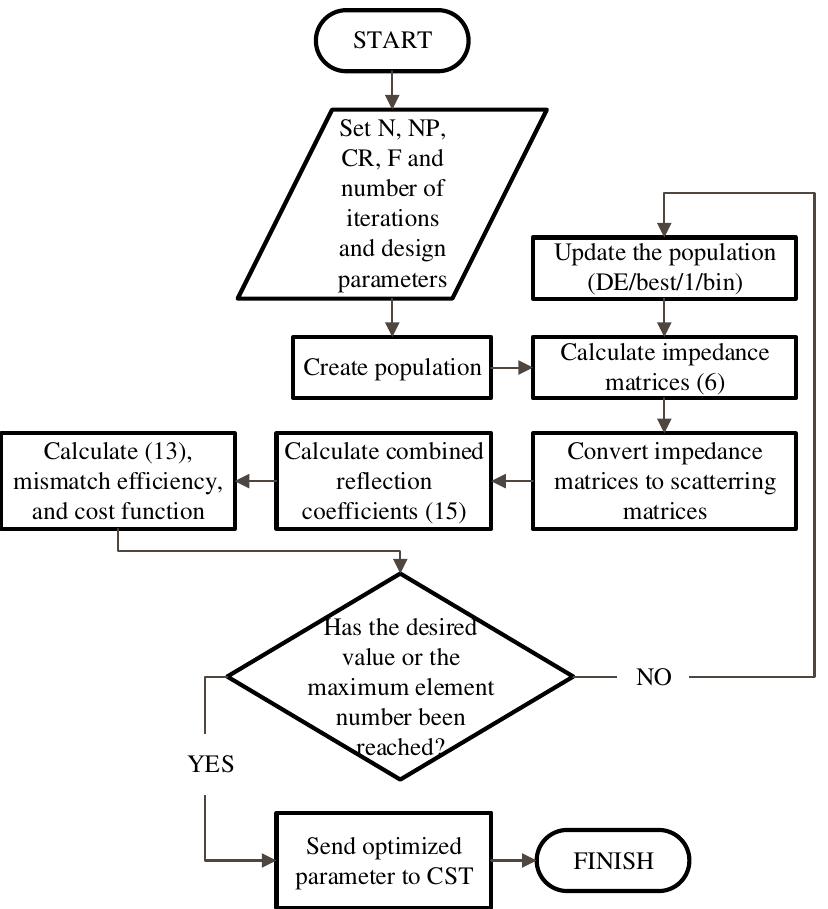}
    \caption{Flowchart of the proposed algorithm to design a $N$-element dipole unit array with high realized gain. Here, $N$ refers to the number of elements spanning from two to ten).}
    \label{fig:algorithm}
\end{figure}
The optimization process, as depicted in Fig. \ref{fig:algorithm}, is presented as a visual flow chart. It begins with a two-element array configured with the predefined parameters ($NP, CR, F$), the number of iterations, and randomly assigned the design attributes, including the currents, lengths, and distances.

The input impedance matrix in (\ref{Prad}) is computed for each member of the population, following the methodology described in \cite{Orfanidis}. Subsequently, the scattering matrix is obtained by applying the relevant transformation formulas, as detailed in \cite{Pozar}. Using this scattering matrix, the reflection coefficients for the combined active elements are calculated. More specifically under simultaneous excitation, we have \cite{MultiCharacteristic}:
\begin{equation}
\Gamma_\mathrm{nc} = \frac{1}{I_\mathrm{n}}\sum_\mathrm{n,m} \Gamma_\mathrm{nm}I_\mathrm{m},
\end{equation}
where, the reflection coefficients of the combined active elements are represented by $\Gamma_\mathrm{nc}$. At each iteration, the algorithm checks if the desired target has been attained; if not, the "$DE/best/1/bin$" algorithm updates the population and continues the process.
\begin{figure}[h]
    \centering
    \includegraphics[width=1\linewidth]{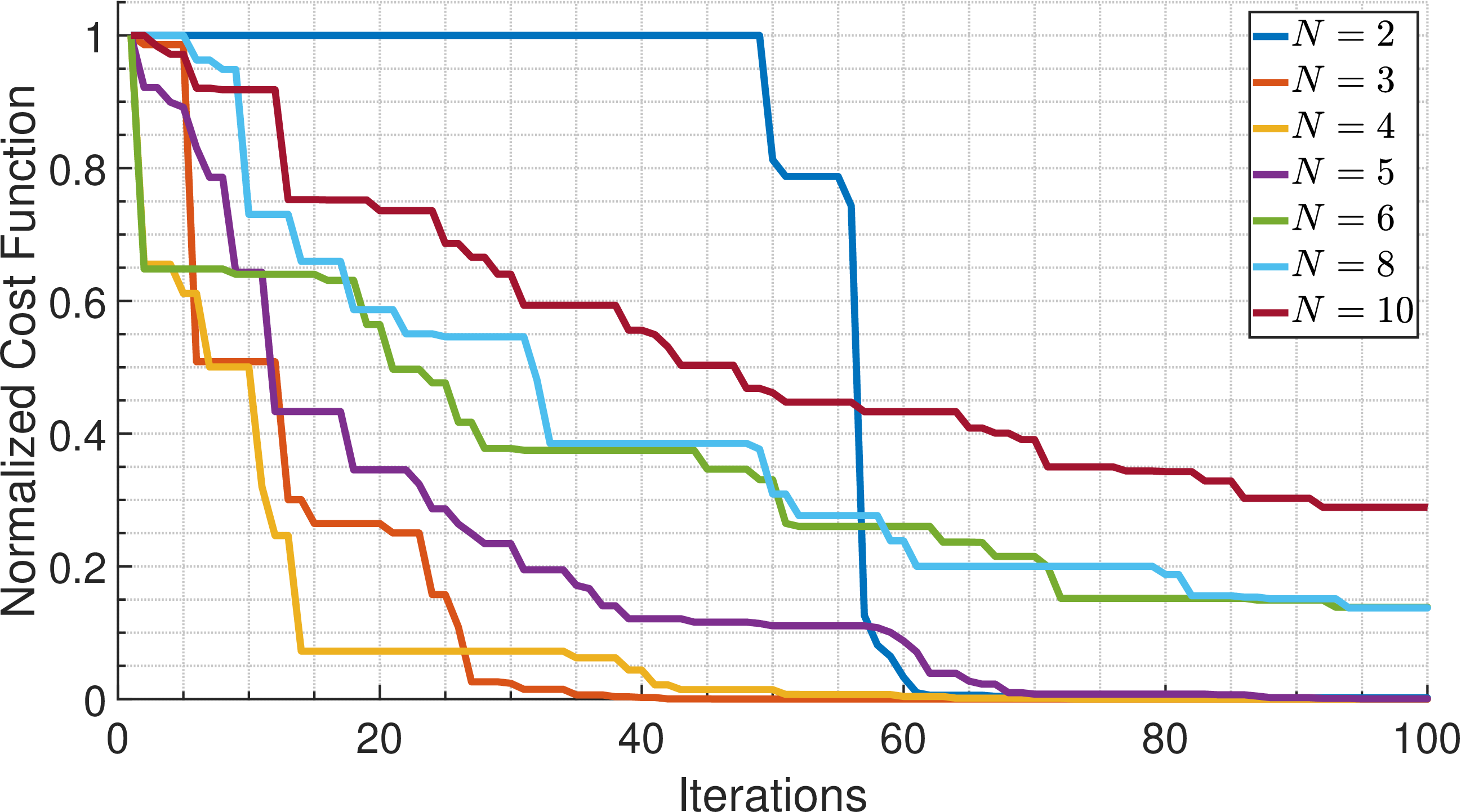}
    \caption{Convergence of the cost function (\ref{eq:cost}).}
    \label{fig:cost}
\end{figure}

\begin{table*}[h]
    \centering
    \caption{The real parts of the input impedances (in $\Omega$) of the optimized array configurations.}
    \begin{tabular}{l|c|c|c|c|c|c|c|c|c|}
    \hline
    \hline
    No. & $N=2$ & $N=3$ & $N=4$ & $N=5$ & $N=6$ & $N=7$ & $N=8$ & $N=9$ & $N=10$ \\
    \hline
    1 & 41.62 & 39.16 & 45.24 & 32.15 & 26.23 & 38.65 & 36.01 & 31.57 & 11.29 \\
    \hline
    2 & 18.92 & 33.00 & 32.50 & 30.60 & 33.48 & 21.47 & 20.61 & 26.99 & 31.12 \\
    \hline
    3 & - & 20.06 & 33.12 & 41.12 & 43.97 & 31.25 & 24.42 & 17.63 & 29.17 \\
    \hline
    4 & - & - & 16.40 & 28.78 & 39.60 & 21.20 & 33.52 & 51.24 & 28.37 \\
    \hline
    5 & - & - & - & 17.69 & 29.99 & 92.36 & 25.85 & -82.25 & 56.28 \\
    \hline
    6 & - & - & - & - & 28.23 & 30.68 & -76.33 & 66.75 & -54.52 \\
    \hline
    7 & - & - & - & - & - & 18.63 & 31.95 & 34.87 & 27.61 \\
    \hline
    8 & - & - & - & - & - & - & 36.83 & 27.74 & 138.85 \\
    \hline
    9 & - & - & - & - & - & - & - & -66.30 & 17.40 \\
    \hline
    10 & - & - & - & - & - & - & - & - & -72.27 \\
    \hline
    \end{tabular}
    \label{tab:input}
\end{table*}

The cost-convergence graph obtained from the optimization process for each configuration is presented in Fig. \ref{fig:cost}. The figure reveals that a maximum of 5 elements is the most meaningful choice for a super-directive antenna array. The cost functions deviate significantly from the target values in configurations with more elements. This divergence is primarily attributed to the increasing dominance of mutual coupling at distances less than half-wavelength when more elements are excited simultaneously. As a result, the input impedances of some elements in the array become too low, sometimes even negative, to be matched. More specifically, as shown in Table \ref{tab:input} obtained by CST, the sixth element of the $8$ element array, the fifth and ninth elements of the $9$ element array, and the sixth and tenth elements of the $10$ element array exhibit negative input impedances. Consequently, achieving super-directivity becomes particularly challenging, especially in arrays of more than five elements. 

\begin{figure}[h]
    \centering
    \includegraphics[width=1\linewidth]{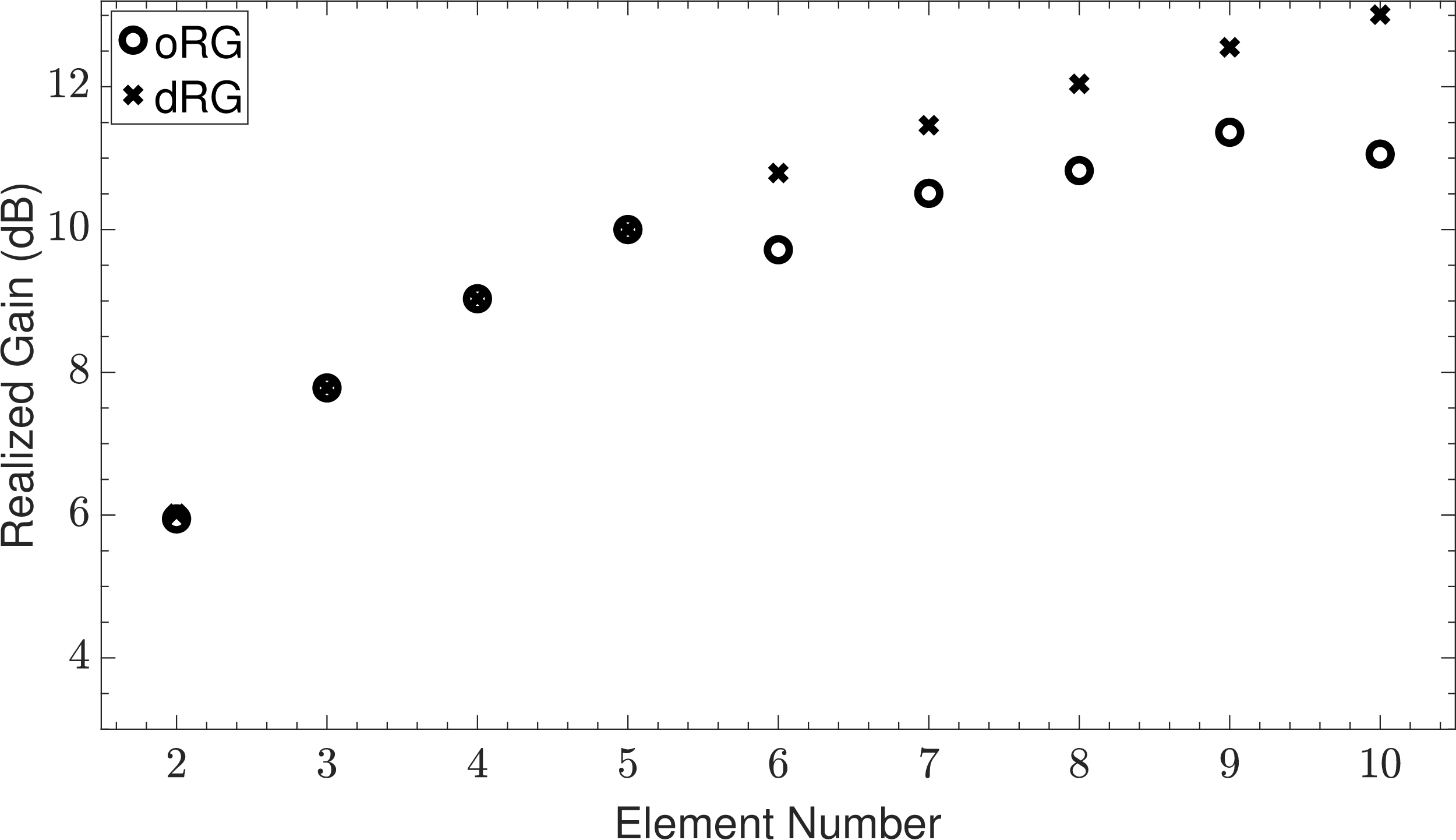}
    \caption{The representation of the difference between the optimized realized gain (oRG) and the desired realized gain (dRG) versus the number of elements obtained by the optimization.}
    \label{fig:oRG_dRG}
\end{figure}

This trend is also evident in Fig. \ref{fig:oRG_dRG}, illustrating the desired (dRG) and optimized realized gains (oRG). Optimizing configurations with up to five elements achieves the target values successfully. However, as the number of elements increases, the difference between the targeted and optimized realized gains significantly widens, mainly due to the amplifying mismatch effect.

\section{Numerical Results and Discussion}
\label{sec:Num}
This section presents the numerical results after determining the super-directive array configuration that maximizes the number of elements achievable through mutual coupling. First, the parameters derived from the optimization process are presented, followed by a comparative analysis of the realized gain versus azimuth angle achieved using MATLAB and CST programs. Subsequently, we compare the realized gain and total efficiency (including radiation and mismatch efficiency) obtained using CST  under various configurations, highlighting the benefits of the proposed super-directive array.
\begin{table}[h]
    \centering
       \caption{Ideal positions (in mm) of the unit array elements achieved from the optimization process.}
        \begin{tabular}{l|c|c|c|c|c|}
        \hline
        \hline
       No. &  Position (mm)  & Amplitude & Phase ($^\circ$) & Length ($\lambda$) \\
         \hline
       1 &-23.1083 & 0.9821  & 46.09 & 0.443 \\
                      \hline
               2 &-13.6325 &  1  & -166.94 & 0.444\\
              \hline
         3 &-0.0505 &  1 & 2.48 & 0.439 \\
              \hline
        4 & 12.7182 & 0.9794 & 159.31 & 0.453 \\
        \hline
         5 & 23.1083 & 0.7805 & -48.95 & 0.481   \\
        \hline
    \end{tabular}
    \label{tab:my_label}
\end{table}

The parameters obtained from the optimization step, including the excitation amplitudes, phases, positions on the $x$-axis, and antenna lengths, for the maximum element super-directive array topology are provided in Table \ref{tab:my_label}. The resulting five-element array antenna occupies $29.87\%$ less space than the uncoupled array with elements positioned at a conventional half-wavelength distance. In comparison to the conventional uncoupled array configuration, the distance between each successive element ($d_{\mathrm{n,m}}$, where $\mathrm{n,m}$ are consecutive integers) is less than half a wavelength (more specifically, $d_{\mathrm{1,2}}=0.31\lambda, d_{\mathrm{2,3}}=0.45\lambda, d_{\mathrm{3,4}}=0.42\lambda$, and $d_{\mathrm{4,5}}=0.34\lambda$). Regarding the optimized excitation amplitudes, although the excitation amplitudes of the first four elements are relatively close to each other, the excitation amplitude of the fifth element is
approximately $20\%$ lower.

\begin{figure}[h]
    \centering
    \includegraphics[width=1\linewidth]{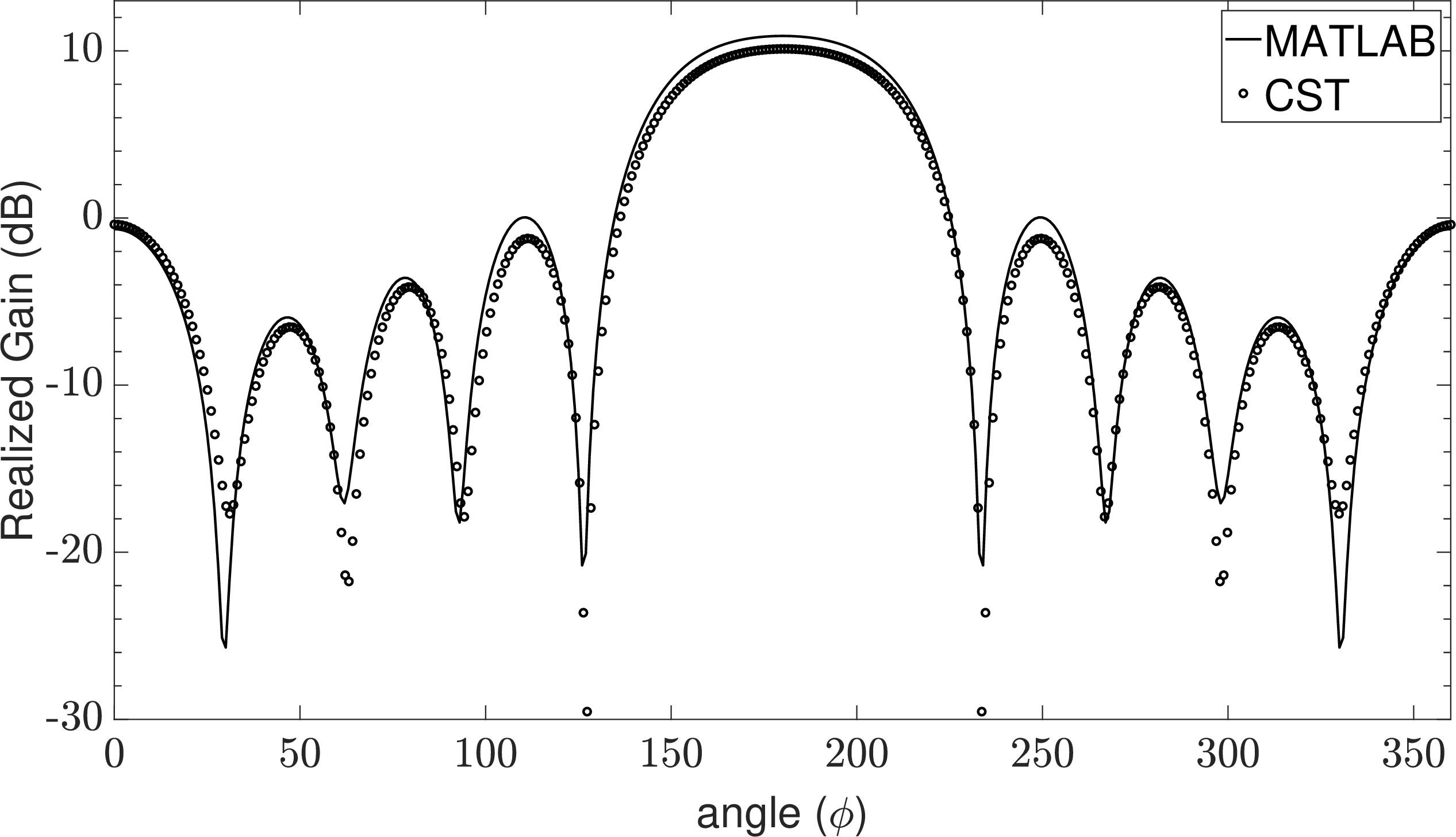}
    \caption{MATLAB and CST comparison of the realized gain results obtained for the optimized ${N=5}$ array.}
    \label{fig:Realizedgainfastest}
\end{figure}

Figure \ref{fig:Realizedgainfastest} shows the comparison of MATLAB and CST results of the realized gain graph of the five-element array. As can be seen from the graph, the optimization results and full-wave simulation results are very close to each other. The small differences can be attributed to the full-wave simulation performing a more comprehensive simulation of the array performance by taking into account all material properties.
\begin{table}[h]
    \centering
       \caption{Comparison table of the optimized super-directive array with the maximum elements ($N=5$) to different configurations.}
        \begin{tabular}{l|c|c|c|c|c|}
        \hline
        \hline
       Conf. & $\#1$ &  $\#2$ & ULA &  Th. Exc.\cite{OptCurrent}& Optimized \\
         \hline
       R. G. (dB) &8.54 & 6.85 & $9.90 \footnotemark$   &7.89 &10.10 \\
                      \hline
              Tot. Eff. ($\%$) & 55.34 & 80.64& 94.13 & 43.39 & 81.60\\
              
        \hline
    \end{tabular}
    \label{tab:comparison}
\end{table}
\footnotetext{Recall that in classical ULA, the absence of phase differences between elements results in maximum radiation in the broadside direction, while super-directive arrays exhibit maximum radiation towards the end-fire.}

After numerical validation, we investigated three different array configurations, as outlined in Table \ref{tab:comparison}, to highlight the benefits of the optimized setup. In the first scenario, referred to as $\#1$, we kept the elements' lengths constant at half-wavelength, in contrast to the configured array. Notably, the realized gain exceeded that of $\#1$ by approximately $44.25\%$. Moving on to $\#2$, where we set the inter-element spacing and elements' lengths uniform as half-wavelength while keeping the other parameters the same in Table \ref{tab:my_label}, the optimized array achieved a nearly $212.23\%$ higher realized gain. 

In our comparison, the third configuration considers a uniform linear array (ULA) with half-wavelength elements and inter-element distances. Here, the optimized array yields a realized gain of approximately $5.31\%$ higher. It is important to note that traditional ULA arrays exhibit the maximum radiation in the broadside direction, which differs from super-directive arrays known for their maximum radiation in the end-fire direction. In this context, when we apply the required phase adjustments to align the ULA towards the end fire, the resulting gain is $6.30$ dB, surpassing the ULA's realized gain by more than $4$ dB.
Moreover, when we maintain the same overall array size along the $x$-axis as the optimized array, introduce a uniform distance of approximately $0.37 \lambda$ between elements, and apply known current values from existing literature for array excitation \cite{OptCurrent}, the optimized array achieves approximately $67.20\%$ higher realized gain.

\section{Conclusion}
\label{sec:Conclusion}
In this study, we examined the role of the number of elements for the design of super-directive array architectures. Our investigation reveals that for a super-directive array with strongly coupled elements, achieving an impedance match to the feed line becomes a challenging task as the number of elements increases. Consequently, this poses a multi-dimensional optimization problem involving an exhaustive assessment of several design parameters, such as the element positions, the element dimensions, and the element excitation amplitudes and phases. Thus, establishing a constraint on the upper bound limit of the number of elements is of significant importance for the design of super-directive arrays.

To address this challenge, we proposed a DE-based optimization algorithm. In particular, we investigated the upper-bound limit on the number of array elements for the optimization of the realized gain parameter. As a result of the optimization, it was observed that super-directivity can be achieved in an array with a maximum of five elements. This array was then analyzed in a full-wave simulation program to verify the algorithm's accuracy.

Our findings showcased the noteworthy characteristics of the suggested super-directive array topology, demonstrating impressive levels of gain ($10.10$ dB) and total efficiency ($81.60\%$). This positions it as a promising contender for densely packed array uses, especially in applications involving massive MIMO. Our findings underscore the possibility of significant enhancements in highly focused arrays by tackling discrepancies and refining array configuration through the utilization of the DE (or other optimization) algorithm.

\bibliographystyle{IEEEtran} 				
\bibliography{referans.bib}
\end{document}